\documentclass[reprint]{JASA-EL}
\usepackage{siunitx}
\usepackage{setspace} 

\begin{document}

\title[JASA-EL/B-CLEAN-SC]{B-CLEAN-SC: CLEAN-SC for broadband sources}
\author{Armin Goudarzi}
\email{armin.goudarzi@dlr.de}
\correspondingauthor
\affiliation{German Aerospace Center (DLR), 37073 G\"ottingen, Germany}
\date{\today} 
\preprint{A. Goudarzi, JASA-EL}  

\singlespacing

\begin{abstract}
This paper presents B-CLEAN-SC, a variation of CLEAN-SC for broadband sources. Opposed to CLEAN-SC, which ``deconvolves'' the beamforming map for each frequency individually, B-CLEAN-SC processes frequency intervals. Instead of performing a deconvolution iteration at the location of the maximum level, B-CLEAN-SC performs it at the location of the over-frequency-averaged maximum to improve the location estimation. The method is validated and compared to standard CLEAN-SC on synthetic cases, and real-world experiments, for broad- and narrowband sources. It improves the source reconstruction at low and high frequencies and suppresses noise, while it only increases the need for memory but not computational effort.
\end{abstract}
\maketitle


\section{\label{sec:BCLEAN:Introduction}Introduction}
Conventional beamforming is a well-established tool to identify and quantify sound sources on complex objects, such as cars, trains, and \replaced[R2C10]{planes}{aircrafts}~\citep{MerinoMartinez2019a}. Naive methods estimate the sound power by virtually steering the Cross Spectral Matrix (CSM) to different focus points to obtain an independent estimation for each focus point. The resulting beamforming map is convoluted with the array's Point Spread Function (PSF), which limits the resolution at low frequencies by the array's aperture and at high frequencies by aliasing that results from the discrete microphone spacing. More advanced methods exist, such as gridless methods~\citep{Kujawski2022,Sarradj2022,Chardon2023, Goudarzi2023}. However, they are computationally expensive and often only proven to work on academic examples.

There exist a variety of ``deconvolution'' methods that aim in reconstructing the true source distribution from the so-called dirty beamforming maps. While advanced source reconstruction methods such as DAMAS~\citep{Brooks2006,Ehrenfried2007,Chardon2021b} exist, CLEAN-SC~\citep{Sijtsma2007} is the gold standard in industrial environments~\citep{Ahlefeldt2016,Ahlefeldt2023}, because it is \deleted[R2C11]{extremely }fast and robust. 

CLEAN-SC solves the deconvolution iteratively at each individual frequency. It assumes a dominant source \added[R2C12]{per iteration} so that the dirty map is dominated by its PSF. It then estimates that the source is located at the location of maximum Power Spectral Density (PSD) in the map and measures the coherence between the location and all other locations. It then subtracts the source from the \replaced{Cross-Spectral Matrix (CSM)}{CSM} and dirty map. It then repeats the process to find \replaced[R2C13]{the second source and so on}{additional sources} until a stopping criterion is met. This process works \deleted[R2C14]{extremely}well \added[R2C14]{for spatially compact sources~\citep{MerinoMartinez2020}} at medium frequencies, where the PSF shows pronounced main-lobes and low side-lobes. At \deleted{very}low frequencies (compared to the array's aperture) the PSF of two adjacent sources will overlap and form a single blob in the dirty map. Thus, the maximum of the dirty map is no longer located at a true source position, but between multiple source positions. At these low frequencies, CLEAN-SC fails to identify the true sources and reconstructs the PSD wrongly. At \deleted{very}high frequencies the focus grid can often no longer resolve the main-lobe. Additionally, grating-lobes are present in the dirty map which are of the same magnitude as the main-lobe. Thus, the maximum is often positioned at a grating-lobe which results in \deleted{very}noisy CLEAN-SC maps at these high frequencies. The improved algorithm HR-CLEAN-SC~\citep{Sijtsma2017} exists that aims to solve the low-frequency issues of CLEAN-SC, which requires an initial CLEAN-SC solution and an additional iteration to obtain a solution. The spatial resolution of HR-CLEAN-SC is approximately doubled compared to CLEAN-SC, but less so if diagonal removal is applied.

Recently, a variation of the gridless CSM-fitting method Global Optimization (GO) was introduced for broadband sources~\citep{Malgoezar2017,Goudarzi2023} based on the observation, that sources typically have a constant location over frequency~\citep{Goudarzi2021}. Broadband GO showed, that introducing the condition of a shared location over frequency smoothes out local minima in the optimization cost function, which are caused by the side- and grating-lobes of the array's PSF. While the results were superior compared to CLEAN-SC and standard GO, the computational effort makes the method currently not suitable for industry applications~\citep{Goudarzi2023,Chardon2023}.

\clearpage
This paper introduces Broadband-CLEAN-SC (B-CLEAN-SC) which aims to relax the problems of CLEAN-SC at high and low frequencies by adapting the idea of broadband GO: The processing of multiple frequencies at once, so that the side-lobes cancel out, and true source positions can be identified. This is done by introducing a simple change to the CLEAN-SC algorithm: Instead of processing each frequency individually, B-CLEAN-SC processes frequency intervals at once (but still obtains smallband solutions). Here, the only difference lies in the determination of the location, from which the source power is sampled. B-CLEAN-SC averages the dirty maps over the frequency interval and uses the location of the maximum averaged source power. It then performs a standard CLEAN-SC iteration for each of the frequencies in the interval with individual source powers per frequency but at the shared location. Thus, the reconstruction at lower frequencies benefits from the resolution at higher frequencies, and the averaging of side- and grating-lobes \replaced{stabilized}{stabilizes} the process at \deleted{very}high frequencies.


\section{\label{sec:BCLEAN:Methodology}Methodology}
This Section presents the standard CLEAN-SC algorithm, and the proposed B-CLEAN-SC algorithm. 

\subsection{\label{sec:BCLEAN:standardCLEAN}Standard CLEAN-SC}
CLEAN-SC is based on the idea that the coherence $\Gamma_{jk}^2$ between an arbitrary focus point $\mathbf{x}_k$ and all other focus points $\mathbf{x}_j$ can be estimated by steering the CSM to the focus points with

\begin{equation}\label{eq:BCLEAN:coherence}
\Gamma_{jk}^2 = \frac{|\mathbf{w}^*_j \mathbf{C}\mathbf{w}_k|^2}{(\mathbf{w}^*_j \mathbf{C}\mathbf{w}_j)(\mathbf{w}^*_k \mathbf{C}\mathbf{w}_k)} = \frac{|A_{jk}|^2}{A_{jj}A_{kk}}\,,
\end{equation}
where $\mathbf{w}$ is an arbitrary steering vector~\citep{Sarradj2012}\added{ and $(\dots)^*$ denotes the Hermetian transpose}. Removing the coherent parts of a source removes the PSF (but also distributed sources) from the map. This is performed iteratively with the Algorithm~\ref{alg:BCLEAN:CLEANSC}, where $n$ is the current iteration, for a maximum number of $N$ iterations, or until a stopping criterion is met~\citep{Sijtsma2007}\replaced{.}{,} $f\in \mathbf{f}$ is the current frequency\added[R2C18]{, $\mathbf{f}=[f_1,f2_,\dots]$ is a frequency vector}, $A$ is the conventional beamforming result for the steering vector $\mathbf{w}$, \deleted{and }$\mathbf{x}$ is a list of all focus points\replaced{.}{,} $\mathbf{C}$ is the dirty CSM, $\mathbf{G}$ is the CSM of the iteratively identified source, and $\mathbf{Q}$ is the final CLEAN-SC estimation of the ``deconvolved'' map. For stability, a loop gain $0< \alpha\le1$ is used. \replaced[R1C01]{For convenience the algorithm is described without Diagonal Removal (DR)~\hbox{\citep{Sijtsma2007}}, since it only adds an identical step to both CLEAN-SC and B-CLEAN-SC.}{CLEAN-SC can be performed with Diagonal Removal (DR) by iteratively adjusting the steering vectors, where $\mathbf{I}$ is the identity matrix and $\circ$ is the Hadamard product in the algorithm.}

\begin{algorithm}[h]
\baselineskip=10pt
\caption{\label{alg:BCLEAN:CLEANSC}Standard CLEAN-SC.}
\begin{algorithmic}
\STATE \textbf{FUNCTION CLEAN-SC}$(\mathbf{C},\mathbf{w},\alpha)$:
\STATE $\mathbf{Q}(\mathbf{f},\mathbf{x}) \gets  \mathbf{0}$
\FOR {$f$ in $\mathbf{f}$}
    \STATE $n \gets 0$
    \STATE $\mathbf{A}_{jj}\gets\mathbf{w}_j^*(f)\mathbf{C}(f)\mathbf{w}_j(f)$
    \WHILE {$n\le N$} \quad\COMMENT{or an other arbitrary stopping criterion is met}
        \STATE $n \gets n+1$
        \STATE $k \gets \text{argmax}_j(\mathbf{A}_{jj})$
        \STATE $A_{kk} \gets \mathbf{A}_{jj}(\mathbf{x}_k)$ \quad\COMMENT{find pos. of max. amplitude}
        \STATE $\mathbf{h}\gets \frac{\mathbf{C}(f)\mathbf{w}_k(f)}{A_{kk}}$ \quad\COMMENT{find steering vector to the corresp. loc.}
        \IF{DR} 
        \STATE $\mathbf{H}\gets \mathbf{I}\circ \mathbf{h}\mathbf{h}^*$ \quad\COMMENT{diag. matrix from steering vector}
        \STATE $\mathbf{h}\gets\frac{1}{1+\mathbf{w}^*\mathbf{H}\mathbf{w}}\left(\frac{\mathbf{C}(f)\mathbf{w}}{\mathbf{w}^*\mathbf{C}(f)\mathbf{w}}+\mathbf{H}\mathbf{w}\right)$ \quad\COMMENT{iteratively find steering vector if DR}
        \ENDIF
        
        \STATE $\mathbf{G}\gets A_{kk} \mathbf{h}\mathbf{h}^*$ \quad\COMMENT{calc. CSM for the identified source}
        \STATE $\mathbf{C}(f) \gets \mathbf{C}(f)-\alpha\mathbf{G}$ \quad\COMMENT{subtract identified source from dirty CSM}
        \STATE $\mathbf{A}_{jj}\gets  \mathbf{A}_{jj} - \alpha\mathbf{w}_j^*(f)\mathbf{G}(f)\mathbf{w}_j(f)$ \quad\COMMENT{subtract corresponding beamforming result from dirty map}
        \STATE $\mathbf{Q}(f,\mathbf{x}_k) \gets \mathbf{Q}(f,\mathbf{x}_k) + \alpha A_{kk}$ \quad\COMMENT{add identified source strength to CLEAN-SC output}
    \ENDWHILE
\ENDFOR
\STATE \textbf{return} $\mathbf{Q}(\mathbf{f},\mathbf{x})$
\end{algorithmic}
\end{algorithm}

\subsection{\label{sec:BCLEAN:BCLEAN}B-CLEAN-SC} 
The B-CLEAN-SC algorithm is nearly identical to the CLEAN-SC algorithm, when CLEAN-SC is performed for all frequencies in parallel with the exception, that B-CLEAN-SC performs each iteration $n$ at a shared location $\mathbf{x}_k$ for all frequencies (within the processed interval $\mathbf{f}$). To determine the location, instead of using the maximum of the dirty map $\mathbf{A}_{jj}(f)$ separately for each frequency, the maximum of the over frequency averaged dirty map is used
\begin{equation}\label{eq:BCLEAN:max_focus_point}
    k = \text{argmax}_j\left(\left\langle \frac{\mathbf{A}_{ijj}}{\text{max}_j(\mathbf{A}_{ijj}^0)}\right\rangle_i\right) \,.
\end{equation}
Here, $\mathbf{A}_{ijj}^0$ denotes the original dirty map prior to subtractions. $i$ denotes the index of the frequency $f_i\in\mathbf{f}$, $j$ denotes the index of the focus point $\mathbf{x}_j$. The subscript of the average operator $\langle\dots\rangle$ or the maximum argument operator indicates the dimension over which they are applied. $\mathbf{A}_{ijj}^0$ is an estimation for the frequency-dependent amplitude of the overall source power (which typically decreases over frequency for aeroacoustic sources). The normalization by its maximum compensates for this behavior. Eq.~\ref{eq:BCLEAN:max_focus_point} is the only addition to the CLEAN-SC algorithm to obtain B-CLEAN-SC, see Algorithm~\ref{alg:BCLEAN:BCLEANSC}. The algorithm is given for a frequency interval $\mathbf{f}$, if the frequency interval does not cover the full frequency range, B-CLEAN-SC \replaced{can be}{is} performed sequentially for multiple intervals.

\begin{algorithm}[h]
\baselineskip=10pt
\caption{\label{alg:BCLEAN:BCLEANSC}B-CLEAN-SC for a frequency interval $\mathbf{f}$.}
\begin{algorithmic}
\STATE \textbf{FUNCTION B-CLEAN-SC}$(\mathbf{C},\mathbf{w},\alpha)$:
\STATE $\mathbf{Q} \gets  \mathbf{0}$
\STATE $n \gets 0$
\STATE $\mathbf{A}_{ijj}^0\gets\mathbf{w}_{ij}^*\mathbf{C}_i\mathbf{w}_{ij}$
\WHILE {$n\le N$}
    \STATE $n \gets n+1$
    \STATE $\mathbf{\hat{A}}_{jj}\gets \left\langle \frac{\mathbf{A}_{ijj}}{\text{max}_j(\mathbf{A}_{ijj}^0)}\right\rangle_i$
    \STATE $k \gets \text{argmax}_j(\hat{A}_{jj})$ \quad\COMMENT{change to the CLEAN-SC algorithm}
    \STATE $A_{ikk} \gets \mathbf{A}_{ijj}(\mathbf{x}_k)$
    \STATE $\mathbf{h}_{ik}\gets \frac{\mathbf{C}_i\mathbf{w}_{ik}}{A_{ikk}}$
    \IF{DR} 
        \STATE $\mathbf{H}_{ikk}\gets\mathbf{h}_{ik}\mathbf{h}^*_{ik}\mathbf{I}_{kk}$ 
        \STATE $\mathbf{h}_{ik}\gets\frac{1}{1+\mathbf{w}_{ik}^*\mathbf{H}_{ikk}\mathbf{w}_{ik}}\left(\frac{\mathbf{C}_{i}\mathbf{w}_{ik}}{\mathbf{w}_{ik}^*\mathbf{C}_{i}\mathbf{w}_{ik}}+\mathbf{H}_{ikk}\mathbf{w}_{ik}\right)$ 
    \ENDIF
    \STATE $\mathbf{G}_i\gets A_{ikk} \mathbf{h}_{ik}\mathbf{h}_{ik}^*$
    \STATE $\mathbf{C}_i \gets \mathbf{C}_i-\alpha\mathbf{G}_i$
    \STATE $\mathbf{A}_{ijj}\gets  \mathbf{A}_{ijj} - \alpha\mathbf{w}_{ij}^*\mathbf{G}_i\mathbf{w}_{ij}$
    \STATE $\mathbf{Q}(f_i,\mathbf{x}_k) \gets \mathbf{Q}(f_i,\mathbf{x}_k) + \alpha A_{ikk}$
\ENDWHILE
\STATE \textbf{return} $\mathbf{Q}(\mathbf{f},\mathbf{x})$
\end{algorithmic}
\end{algorithm}

Note, that the position $\mathbf{x}_k$ is not necessarily located on the main-lobe of a dominant source for all frequencies if the sources have a strong frequency-dependent power. Especially at low frequencies, where the PSF of a dominant source may cover all other sources and dominate the estimated power at their true positions, this would lead to an overestimation of their power, and a subtraction of the main-lobe, when subtracting coherent portions of the map~\citep{Sijtsma2017}. To relax this issue, a low gain factor $\alpha$ is needed, so that the number of necessary B-CLEAN-SC iterations increases. Since only the initial calculation of the dirty map is computationally expensive the extra iterations are not performance relevant.

\section{\label{sec:BCLEAN:Results}Results}
This section presents \replaced[R1C02]{the results of four}{three} different cases. Section~\ref{sec:BCLEAN:Synthetic_results} presents \replaced[R1C02]{two}{a} synthetic example\deleted{s} that aim\added{s} to clarify the behavior of CLEAN-SC and B-CLEAN-SC. Section~\ref{sec:BCLEAN:Exp_results_1} presents \replaced[R2C28]{a real}{an open wind tunnel} experiment with ground truth, so that the methods can be evaluated quantitatively. Last, Section~\ref{sec:BCLEAN:Exp_results_2} presents a \replaced[R2C28]{real-world}{closed} wind tunnel experiment without ground truth, based on which the methods are evaluated qualitatively. Throughout this section, CLEAN-SC will be performed with diagonal removal, a maximum of $3N_S$ iterations per frequency where $N_S$ is the number of true sources, and a gain factor of $\alpha=0.9$ per iteration. B-CLEAN-SC will be performed with diagonal removal, a maximum of $10N_S$ iterations, and $\alpha=0.1$ per iteration. To reduce the visual complexity of the results, beamforming maps are obtained only in 1D for case 1, and 2D for cases 2 and 3 with steering vector formulation III~\citep{Sarradj2012}\deleted{, with DR}. \added[R2C24c)]{The results will be presented over the Helmholtz number $\text{He}=fD/a$, where $D$ is the array's aperture, and $a$ is the speed of sound.}

\subsection{\label{sec:BCLEAN:Synthetic_results}Synthetic results} 
\replaced{We propose case 1,}{Case 1 is} a synthetic 1D example that highlights the differences between standard CLEAN-SC and B-CLEAN-SC. The array is located at $-0.5 \le x \le 0.5$, $y=0$. There are three sources $S_i$ at $x_1=0$, $x_2=0.1$, $x_3=0.5$, $y=0.5$. The CSM is calculated at 256 frequencies $f_\text{max}=\SI{8192}{\hertz}$, $\Delta f = \SI{32}{\hertz}$. The focus grid is located at $-1 \le x \le 1$, $y=0.5$, $\Delta x=\SI{0.004}{\metre}$. The PSD of $S_1$ linearly increases over frequency from $\text{PSD}_1(f_0)=\SI{-10}{\decibel}$ to $\text{PSD}_1(f_{256})=\SI{0}{\decibel}$. The PSD of $S_2$ linearly decreases in the same way so that $S_2$ dominates at low frequencies and $S_1$ dominates at high frequencies. Additionally, $S_3$ is a smallband source that is only present at $\SI{3616}{\hertz}\le f \le \SI{3840}{\hertz}$ at $\SI{-10}{\decibel}$. For B-CLEAN-SC, the frequencies are processed in intervals of $\Delta f=\SI{2048}{\hertz}$.

\begin{figure}[h]
\centerline{\includegraphics[width=\linewidth]{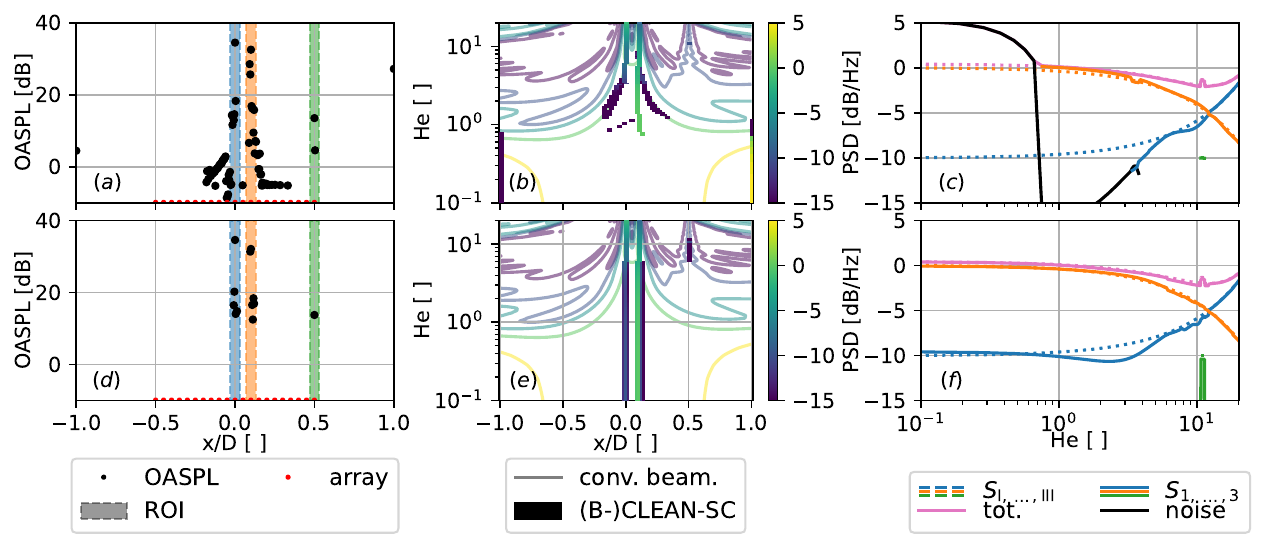}}
\caption{\label{fig:BCLEAN:Figure1}Case 1, the top row shows CLEAN-SC, the bottom row B-CLEAN-SC results. $(a)$\&$(d)$ show the OASPL($x/D$), integrated over all frequencies. The sensor positions are marked at an arbitrary $y$-location. The shaded areas represent the ROI. $(d)$\&$(e)$ show the conventional beamforming result and the sparse (B-)CLEAN-SC result, the color indicates the PSD. $(e)$\&$(f)$ show the resulting spectra, integrated from the ROI in $(a)$\&$(d)$. The sources are depicted with different colors: $S_1$ (blue), $S_2$ (orange), $S_3$ (green), and the total integrated power (magenta). (B-)CLEAN-SC results that are not located within any ROI are spatially integrated and classified as noise (black). The ground truth is depicted with dotted lines and Latin numbers, the ROI estimation with full lines and Arabic numbers.}
\raggedright
\end{figure}

Figure~\ref{fig:BCLEAN:Figure1} shows the results of case 1. \replaced{Figure~\ref{fig:BCLEAN:Figure1} $(a)$ and $(c)$ show the CLEAN-SC results, Figure~\ref{fig:BCLEAN:Figure1} $(b)$ and $(d)$ show the B-CLEAN-SC results. $(a)$ and $(b)$ show the source reconstruction over space and frequency, and the color depicts the PSD. The underlying color-map depicts the conventional result for reference. The vertical colored lines represent Regions Of Interest (ROI) around the true source locations. $(c)$ and $(d)$}{Figure~\ref{fig:BCLEAN:Figure1} $(e)$\&$(f)$} show the \replaced{corresponding}{estimated} PSD, integrated from the same colored ROI \added{in Figure~\ref{fig:BCLEAN:Figure1} $(a)$\&$(d)$}. The black lines \replaced{indicate}{represent} noise, integrated from the area that does not correspond to any ROI \added{indicating beamforming and deconvolution artifacts}. Additionally, a \replaced[R2C24a)]{red}{magenta} line shows the integration of all sources within the map, as an estimation of the overall sound power. The ground truth is depicted with dotted lines for reference. 

CLEAN-SC reconstructs the dominant source $S_2$ well down to \replaced{$f\ge\SI{200}{\hertz}$}{$\text{He}\ge0.8$}, below which the maximum within the dirty map is \replaced{assumed}{estimated} with a wrong level \added{along the side-lobes and then} at the edges of the focal range. For $S_1$ the PSD reconstruction works well down to \replaced{$f\ge\SI{1}{\kilo\hertz}$}{$\text{He}\ge4$}, below which CLEAN-SC gradually underestimates its power and gradually misses the correct location. The smallband source $S_3$ is reconstructed perfectly. B-CLEAN-SC perfectly estimates the sources' locations. The PSDs are reconstructed well throughout the frequency range, except for an underestimation of $S_1$ at \replaced{$f\approx\SI{1}{\kilo\hertz}$}{$\text{He}\approx3$}. For B-CLEAN-SC, there is no noise.

\subsection{\label{sec:BCLEAN:Exp_results_1}Experiment with ground truth} 
Case 2 \replaced{features}{is} a generic \added[R2C28]{open} wind tunnel experiment \added{at Mach $\text{M}=0.06$} with a streamlined monopole speaker, that is moved to three different locations~\citep{Goudarzi2021,Goudarzi2023}\replaced[R2C27]{. The individual CSMs are used to calculate the ground truth at Mach $\text{M}=0$. Then, a measurement at $\text{M}=0.06$ is performed for all three source positions and their CSMs are added to obtain a problem with three sources.}{ with different spectra and known ground truth.} The sources are located at $x_1=-0.05$, $x_2=0.1$, $x_3=0.25$, $y_{1,2,3}=0.1$, $z_{1,2,3}=0$. The array consists of 7x7 equidistantly spaced microphones with $\Delta x = \Delta y = \SI{0.09}{\metre}$, and is located at $z=-0.65$ \added[R2C28]{outside of the flow}. The equidistant 2D focus grid $\Delta x = \Delta y = \SI{0.005}{\metre}$ covers $-0.3\le x,y \le 0.3$ at $z=0$. The sampling rate is $f_s=\SI{65536}{\hertz}$, and $\Delta f=\SI{512}{\hertz}$.

\begin{figure}[h]
\centerline{\includegraphics[width=\linewidth]{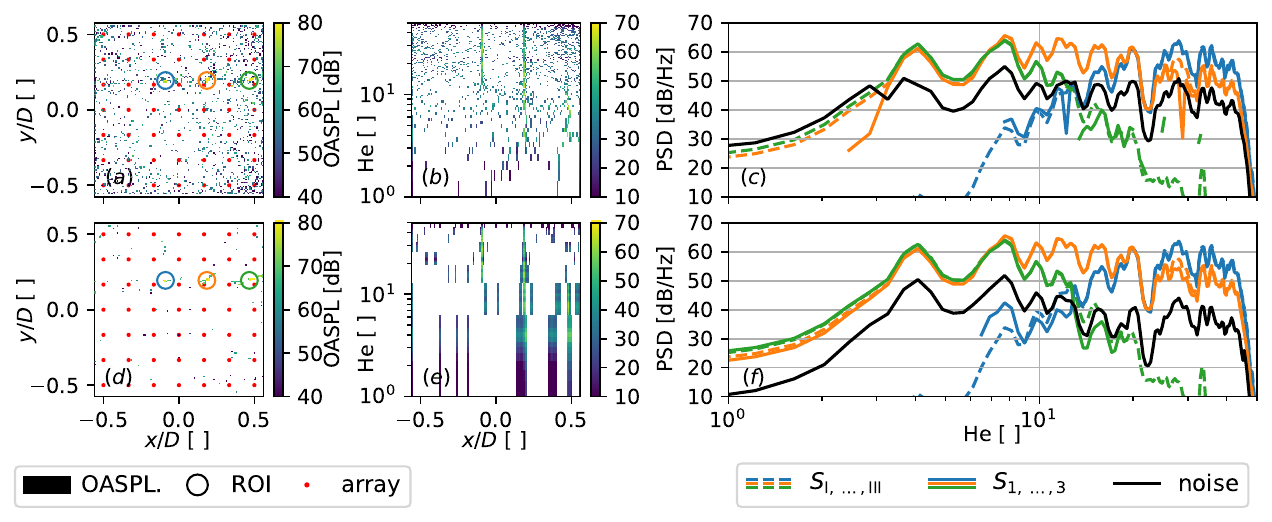}}
\caption{\label{fig:BCLEAN:Figure2}Case 2, the top row shows CLEAN-SC and the bottom row B-CLEAN-SC results. $(a)$\&$(d)$ show the OASPL($x/D,y/D$) with colored ROI centered around the true source locations, $(b)$\&$(e)$ show the PSD($x/D,f$) integrated over the $y$-dimension, $(c)$\&$(f)$ show the ground truth (dotted, Latin numbers) and estimated PSD (solid, Arabic numbers) from the identical colored ROI in $(a)$\&$(d)$. The black line represents noise, integrated from the areas that do not correspond to any ROI.}
\raggedright
\end{figure}

\begin{table}[ht]
\caption{\label{tab:table1}Influence of the frequency interval $\mathbf{f}$ on the resulting error metrics, where $||\mathbf{f}||=128$ are all frequencies. [a] corresponds to the CLEAN-SC result in Figure~\ref{fig:BCLEAN:Figure2}, and [b] to the B-CLEAN-SC result, all with DR, $\alpha=0.1$, $N=30$.}

\begin{ruledtabular}
\begin{tabular}{c|cccccccc}
$||\mathbf{f}||$ & 1\footnotemark[1] & 2 & 4 & 8 & 16 \footnotemark[2] & 32 & 64 & 128\\
\hline
correct PSD~[$\si{\percent}$]& 61.7 & 62.8 & 62.5 & 65.6 & \textbf{66.7} & \textbf{66.7} & 66.1 & 64.8\\
mean error~[$\si{\decibel}$]& 2.3 & 2.2 & 2.5 & 2.5 & \textbf{1.3} & 2.5 & 3.9 & 8.9\\
SNR~[$\si{\decibel}$] & 14.5 & 14.4 & 14.8 & 16.4 & 17.8 & 18.4 & \textbf{22.6} & 19.1\\
\end{tabular}
\end{ruledtabular}
\end{table}

\deleted[R1C04]{Figure~\ref{fig:BCLEAN:Figure2} shows the results for case 2. Figure~\ref{fig:BCLEAN:Figure2} $(a)$ shows CLEAN-SC's spatial source estimation, integrated over the $y$-dimension (to yield a 2D representation). Source $S_1$ (blue), $S_2$ (orange), and $S_3$ (green) are spatially well estimated for $f\ge\SI{3}{\kilo\hertz}$. However, the result is very noisy. Figure~\ref{fig:BCLEAN:Figure2} $(b)$ shows the integrated ROI from the same colored areas in $(a)$. The ROI are circles with a radius of $r=\SI{0.03}{\metre}$ around the true source locations. CLEAN-SC is generally able to estimate the source PSD well. Source $S_1$ is estimated well at high frequencies $f\ge\SI{5}{\kilo\hertz}$. Source $S_2$ is estimated well down to $f \ge \SI{1.5}{\kilo\hertz}$, below which it can no longer be separated from $S_3$. $S_3$, which dominates at low frequencies, is estimated well down to $f \ge \SI{2}{\kilo\hertz}$. Below this frequency, the overall power was estimated well, but could not be attributed to a true source position. Both $S_1$ and $S_3$ are reconstructed down a Signal-to-Signal Ratio (SSR) of around $\text{SSR} = \SI{30}{\decibel}$, which was used as a stopping criterion for CLEAN-SC. Throughout the frequency range, the result is very noisy and the Signal-to-Noise Ratio (SNR) compared to the dominating ROI spectrum is $\text{SNR} = \SI{11.5}{\decibel}$ (averaged over all frequencies, where there exists a ROI spectrum, which is $f\ge\SI{1.5}{\kilo\hertz}$).}

\deleted[R1C04]{Figure~\ref{fig:BCLEAN:Figure2} $(c)$ and $(d)$ show the corresponding B-CLEAN-SC results. For the spatial estimation, the processed frequency intervals with shared source positions are well visible. They result in a sparse positional estimation with less noise. The SNR (compared to the noise outside of ROI) is $\text{SNR}=\SI{17.8}{\decibel}$.}
\deleted[R1C04]{Figure~\ref{fig:BCLEAN:Figure2} shows the results for case 2. Figure~\ref{fig:BCLEAN:Figure2} $(a)$ shows CLEAN-SC's spatial source estimation, integrated over the $y$-dimension (to yield a 2D representation). Source $S_1$ (blue), $S_2$ (orange), and $S_3$ (green) are spatially well estimated for $f\ge\SI{3}{\kilo\hertz}$. However, the result is very noisy. Figure~\ref{fig:BCLEAN:Figure2} $(b)$ shows the integrated ROI from the same colored areas in $(a)$. The ROI are circles with a radius of $r=\SI{0.03}{\metre}$ around the true source locations. CLEAN-SC is generally able to estimate the source PSD well. Source $S_1$ is estimated well at high frequencies $f\ge\SI{5}{\kilo\hertz}$. Source $S_2$ is estimated well down to $f \ge \SI{1.5}{\kilo\hertz}$, below which it can no longer be separated from $S_3$. $S_3$, which dominates at low frequencies, is estimated well down to $f \ge \SI{2}{\kilo\hertz}$. Below this frequency, the overall power was estimated well, but could not be attributed to a true source position. Both $S_1$ and $S_3$ are reconstructed down a Signal-to-Signal Ratio (SSR) of around $\text{SSR} = \SI{30}{\decibel}$, which was used as a stopping criterion for CLEAN-SC. Throughout the frequency range, the result is very noisy and the Signal-to-Noise Ratio (SNR) compared to the dominating ROI spectrum is $\text{SNR} = \SI{11.5}{\decibel}$ (averaged over all frequencies, where there exists a ROI spectrum, which is $f\ge\SI{1.5}{\kilo\hertz}$).}

\deleted[R1C04]{Figure~\ref{fig:BCLEAN:Figure2} $(c)$ and $(d)$ show the corresponding B-CLEAN-SC results. For the spatial estimation, the processed frequency intervals with shared source positions are well visible. They result in a sparse positional estimation with less noise. The SNR (compared to the noise outside of ROI) is $\text{SNR}=\SI{17.8}{\decibel}$.}
\added[R1C04]{Figure~\ref{fig:BCLEAN:Figure2} shows the results for case 2. Figure~\ref{fig:BCLEAN:Figure2} $(a)$\&$(d)$ show that CLEAN-SC results in noisier OASPL maps than B-CLEAN-SC. When integrating the maps over $x$, Figure~\ref{fig:BCLEAN:Figure2} $(b)$\&$(e)$ show that CLEAN-SC is able to determine the correct location down to $\text{He}\approx3$. B-CLEAN-SC correctly determines throughout the frequency range. Strong Side-lobes are reconstructed as ``ghost sources'', that move closer to the true source position with increasing frequency. Figure~\ref{fig:BCLEAN:Figure2} $(c)$\&$(f)$ shows the spectrum estimation. Source $S_1$ is estimated well by CLEAN-SC at $\text{He}\ge8$. Source $S_2$ is estimated well down to $\text{He}\approx1.5$, below which it can no longer be separated from $S_3$, estimated well down to $\text{He}\ge2$. Below this frequency, the overall power was estimated well, but could not be attributed to a true source position, so that it was integrated as noise. Both $S_1$ and $S_3$ are reconstructed down a Signal-to-Signal Ratio (SSR) of around $\text{SSR} = \SI{30}{\decibel}$, which was used as an iteration stopping criterion. B-CLEAN-SC shows similar results, with improved reconstructions of $S_2$ and $S_3$ and lower levels of noise.}

\added[R1C04]{Table~\ref{tab:table1} shows a comparison of three different metrics for exemplary frequency intervals, where $|\dots|$ is the absolute value, $||\dots||$ is the number of elements, and $\langle\dots\rangle_v$ is the average over the variable $v$. The relative frequency interval of the PSD that is correct (within a $\pm\SI{3}{\decibel}$ margin of the ground truth (GT)) is defined for multiple sources $S$ as
\begin{equation}\label{eq:corr_PSD}
    \text{corr. PSD}=\left\langle\frac{|| -3\le|\text{PSD}(S,f)-\text{GT}(S,f)|\le3 ||_f}{||\text{PSD}(S,f)||_f}\right\rangle_S\,,
\end{equation}
the mean absolute error (for frequencies where the PSD is defined, so that $\text{PSD}\neq-\infty~\si{\decibel}$)
\begin{equation}\label{eq:MAE}
    \text{mean error}=\langle\langle|\text{PSD}(f)-\text{GT}(f)|\rangle_f\rangle_S\,,
\end{equation}
and the SNR is the ratio of the maximum source level to the noise level (for frequencies where both are defined)
\begin{equation}\label{eq:SNR}
    \text{SNR}=\langle\underset{S}{\text{max}}(\text{PSD}(S,f))-\text{noise}(f)\rangle_f\,.
\end{equation}
Note, that eq.~\ref{eq:MAE} and eq~\ref{eq:SNR} are conditional errors, since require a defined PSD estimation. Figure~\ref{fig:BCLEAN:Figure2} $(c)$\&$(f)$ show that this is not always the case, so that they must be evaluated as subsidiary results of eq.~\ref{eq:corr_PSD}. E.g., the low frequency CLEAN-SC noise is not captured with the SNR metric, as no signal is present. Table~\ref{tab:table1} shows that B-CLEAN-SC outperforms CLEAN-SC with an increasing frequency interval, with an optimum at $16\le||\mathbf{f}||\le32$. The SNR further improves with an increasing frequency interval which indicates an improvement in the spatial localization, but the spectral estimation deteriorates in return.}

\subsection{\label{sec:BCLEAN:Exp_results_2}Wind tunnel experiment} 
Case 3 is a \added[R2C28]{closed} wind tunnel measurement of a Dornier 728 at $\text{M}=0.125$~\citep{Ahlefeldt2013}. The 2D focus grid $\Delta x = \Delta y = \SI{0.01}{\metre}$ is rotated so that it covers and follows the wing. The spiral array with \added{$D=\SI{1}{\metre}$} consists of 149 microphones\deleted{, has a diameter of approx. $d=\SI{1}{\metre}$}, and is located approx. $\Delta z=\SI{1}{\metre}$ from the wing. The signal is sampled at $f_s=\SI{120}{\kilo\hertz}$ and the CSM is sampled for 128 frequencies at $\Delta f\approx \SI{479}{\hertz}$. Since there exists no ground truth, the results will be only discussed qualitatively.

\begin{figure}[h]
\centerline{\includegraphics[width=\linewidth]{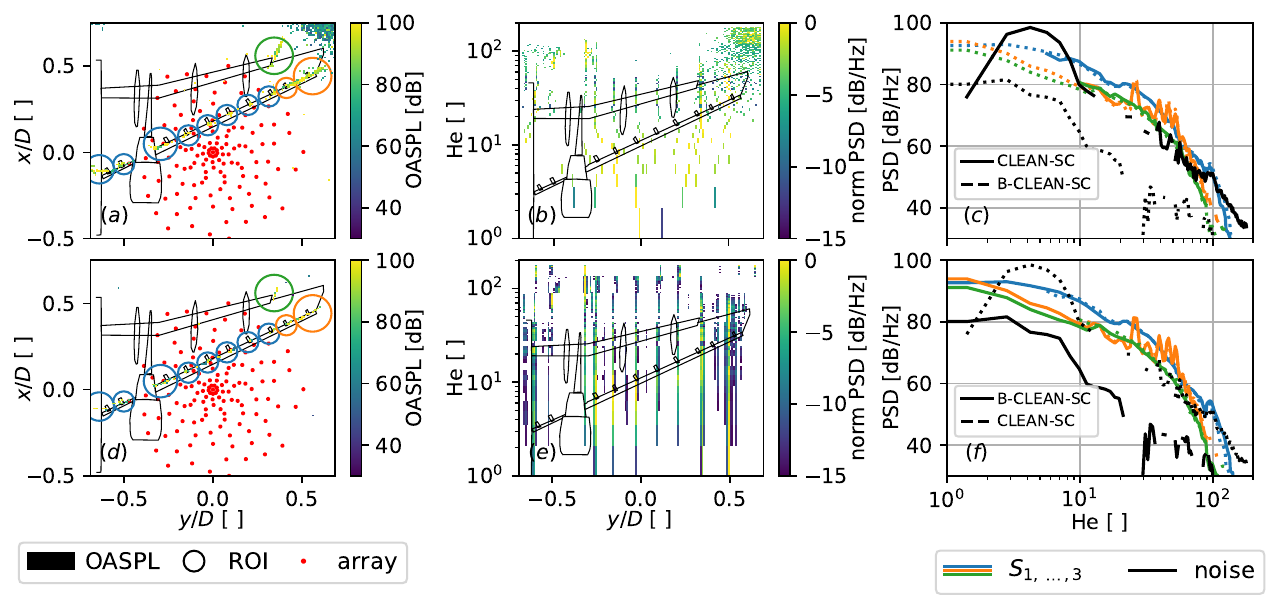}}
\caption{\label{fig:BCLEAN:Figure3}Case 3, the top row shows CLEAN-SC and the bottom row B-CLEAN-SC results. $(a)$\&$(d)$ show the OASPL($x/D,y/D$), $(b)$\&$(e)$ show the normalized PSD($y/D,f$) integrated over the $x$-dimension and normalized per frequency, the model is depicted for reference (but its $x$-information should be ignored). $(c)$\&$(f)$ show the estimated PSD from the identical colored ROI in $(a)$\&$(d)$. While $(c)$ shows the CLEAN-SC results (solid), the B-CLEAN-SC solution is displayed with dotted lines, and in $(f)$ vice versa for comparison. Black lines indicate noise.}
\raggedright
\end{figure}

\replaced{Figure~\ref{fig:BCLEAN:Figure3} $(a)$ and $(b)$ show the estimated source distribution over the $y$-dimension, integrated over $x$. Thus, the only sources that can be confused in this depiction are an outboard slat track and the flap side edge at $y\approx\SI{0.3}{\metre}$. The color-map shows the estimated PSD, normalized per frequency, within a range of $\SI{15}{\decibel}$. The model is depicted for reference. Note, that the $x$-component of the model is plotted, but the color-map does not include any $x$-information. Figure~\ref{fig:BCLEAN:Figure3} $(a)$ shows the CLEAN-SC result and $(b)$ shows the B-CLEAN-SC result. For the CLEAN-SC result one can clearly identify the slat tracks in a frequency range of $\SI{5}{\kilo\hertz} \ge f \ge \SI{15}{\kilo\hertz}$. Above this range, the result mostly shows the inboard Kr\"uger slat, the nacelle area, and the noise for $f\ge\SI{10}{\kilo\hertz}$. Below $f\le\SI{5}{\kilo\hertz}$ the source separation fails. The B-CLEAN-SC result shows the same slat tracks as dominant sources. However, they are also reconstructed at low frequencies $f\le\SI{5}{\kilo\hertz}$ and at high frequencies $f\ge\SI{40}{\kilo\hertz}$. Additionally, there is nearly no noise at high frequencies. Additional sources are located between the sources, which are typically connected to slat cove tones~\hbox{\citep{Goudarzi2022}}. Overall, the location of the estimated sources strongly correlates to the geometrical features of the model and is consistent over the whole frequency range.}{Figure~\ref{fig:BCLEAN:Figure3} $(a)$\&$(d)$ show the estimated OASPL$(x/D,y/D)$. The results correspond well to the geometric features of the wing, but CLEAN-SC shows noise in the top right corner, and sources such as the flap side edge are not well localized. Figure~\ref{fig:BCLEAN:Figure3} $(b)$\&$(e)$ show the PSD$(y,f)$, integrated over $x$ and normalized per frequency. Thus, the only sources that can be confused in this depiction are an outboard slat track and the flap side edge at $y\approx\SI{0.3}{\metre}$. The model is depicted for reference, so that the estimated sources can be attributed to its geometrical features such as the slat tracks. Note, that the $x$-component of the model is plotted along the frequency axis, but the color-map does not include any $x$-information. Within the CLEAN-SC result one can clearly identify slat tracks in a frequency range of $10\le \text{He}\le100$. Otherwise, the result mostly shows the inboard Kr\"uger slat, the nacelle area, and the noise for $y/D\ge0.5$. Below $\text{He}\le10$ the source separation fails. The B-CLEAN-SC result shows the same slat tracks as dominant sources. However, they are also reconstructed at low  and high frequencies. Additionally, there is nearly no noise for $y/D\ge0.5$. Additional sources are visible between the slat tracks, which are typically connected to slat cove tones~\citep{Goudarzi2022}. Overall, the location of the estimated sources strongly correlates to the geometrical features of the model and is consistent over the whole frequency range.}

Based on the \deleted[R2C31]{excessive} analysis of this data~\citep{Goudarzi2022} ROI are defined that cover the inner (Kr\"uger) slat and the slat tracks (blue), the outer slat (orange), and the flap side edge (green). The ROI are chosen, so that the integrated source types are similar~\citep{Goudarzi2022, Ahlefeldt2023}. \replaced{Figure~\ref{fig:BCLEAN:Figure3} $(c)$ shows the ROI, and Figure~\ref{fig:BCLEAN:Figure3} $(d)$ shows the corresponding CLEAN-SC results (dotted) and B-CLEAN-SC results (dashed). Below $f\le\SI{5}{\kilo\hertz}$, CLEAN-SC fails to reconstruct individual sources, as shown in Figure~\ref{fig:BCLEAN:Figure3}, which results in strong noise. B-CLEAN-SC estimates the dominant source to be the slat tracks (which coincides with the overall CLEAN-SC solution), followed by the outer slat and flap side edge. Between $\SI{5}{\kilo\hertz}\ge f\ge\SI{40}{\kilo\hertz}$ the ROI results of both methods are nearly identical. For $f\ge\SI{15}{\kilo\hertz}$ the CLEAN-SC result is contaminated with noise that, based on its spectral shape, originates from the outer slat. For the B-CLEAN-SC result, there is noise throughout the frequency range, however, the SNR is much larger compared to the CLEAN-SC result.}
{Figure~\ref{fig:BCLEAN:Figure3} $(a)$\&$(d)$ show the (identical) ROI, and Figure~\ref{fig:BCLEAN:Figure3} $(c)$\&$(f)$ show the corresponding PSD. Below $\text{He}\le10$, CLEAN-SC fails to reconstruct individual sources, which results in strong noise, additional to the noise $\text{He}\ge40$. B-CLEAN-SC reconstructs the PSD throughout the frequency range with approx. $\SI{20}{\decibel}$ less noise. For frequencies where both methods produce a source spectrum they coincide.
}

\section{\label{sec:BCLEAN:Discussion}Discussion}
Case 1 showed that CLEAN-SC can predict arbitrary results at low frequencies. B-CLEAN-SC fixed this by averaging frequency intervals of dirty maps to determine source locations. This works, as the locations of side- and grating-lobes change with frequency so that they cancel out during the averaging. Additionally, the source location at low frequencies below the Rayleigh resolution \added[R2C32]{limit} is determined based on higher frequencies, where the source positions can still be resolved. The case showed that B-CLEAN-SC also works for sources with a frequency-dependent spectrum and smallband sources. Here, the initial source marker is not guaranteed to be located on the dominant source for all frequencies. Thus, B-CLEAN-SC is prone to ``confuse'' the power contribution of these sources. To relax this problem, a low iteration gain factor of $\alpha=0.1$ was used. Additionally, using frequency intervals instead of using the whole spectrum further relaxes this issue.

Case 2 showed \replaced{how CLEAN-SC and B-CLEAN-SC perform on a generic wind tunnel measurement, featuring a monopole speaker with a ground truth. Overall, both methods performed similarly with two main differences. First,}{that} B-CLEAN-SC \replaced{was}{is} able to correctly determine the location and power of the sources at low frequencies\replaced{. Second,}{ in an open wind tunnel experiment and} its overall noise level was $\SI{6}{\decibel}$ lower compared to CLEAN-SC. \added[R1C04,R2C34]{The introduced metrics and Table~\ref{tab:table1} showed that B-CLEAN-SC improves with increasing frequency intervals in spatial and spectral accuracy up to an optimum at 1/4 of the total frequencies, after which spatial accuracy is traded for a deteriorating spectral estimation. One can possibly account for this by defining frequency-dependent intervals so that the intervals are large at low and high frequencies and small at medium frequencies where CLEAN-SC works well. A lower gain factor further relaxes this issue but increases the number of iterations.}

Case 3 showed \replaced{the performance of both methods on}{that for} a real-world wind tunnel measurement of a Do728\deleted{. Again,} B-CLEAN-SC was able to reconstruct sources throughout the frequency range, compared to CLEAN-SC which identified sources mainly at \replaced[R2C33]{$\SI{5}{\kilo\hertz}\ge f \ge \SI{45}{\kilo\hertz}$}{$10\le \text{He} \le 100$}. Since their location is roughly constant over frequency and corresponds to the geometric features (slat track, flap side edge, etc.) \deleted{we assume }these identified locations \replaced{to be}{are presumably} correct. The B-CLEAN-SC result is less noisy compared to the CLEAN-SC result. The \replaced{source type-dependent ROI integration}{ROI PSD} showed nearly identical results for both methods in the frequency region where CLEAN-SC correctly identified sources\added{, which was smaller compared to B-CLEAN-SC}.

\deleted{For B-CLEAN-SC, the frequency interval has an impact on the results (not shown within the scope of this paper). With an increasing frequency interval, the spatial source estimation is refined, but the spectral estimation gets worse if the dominance of sources strongly varies within the interval. One can possibly account for this behavior by defining frequency-dependent intervals so that the intervals are large at very low and high frequencies and small at medium frequencies where CLEAN-SC works well. A low gain factor relaxes this issue but increases the number of iterations.}

\section{\label{sec:BCLEAN:conclusion}Conclusion}
This paper presented Broadband-CLEAN-SC (B-CLEAN-SC), a variation of CLEAN-SC \replaced[R2C09]{for}{specifically tailored to} broadband sources. B-CLEAN-SC assumes that the location of broadband sources is constant over frequency intervals. For synthetic and experimental wind tunnel data B-CLEAN-SC outperformed CLEAN-SC at low frequencies. For experimental real data, B-CLEAN-SC also resulted in \replaced{$\SI{6}{\decibel}$ less noise}{$\SI{3}{\decibel}$ less broadband noise} throughout the frequency range. On wind tunnel data of a Dornier 728 both methods showed that the source location assumption is valid, improves the spatial estimation of sources, and reduces noise.

The algorithmic difference between CLEAN-SC and B-CLEAN-SC is small. B-CLEAN-SC processes multiple frequencies at once and uses one additional operation per iteration compared to CLEAN-SC. As it requires a lower gain factor, \replaced[R1C04]{more iterations are necessary}{the number of iterations increase inverse proportionally to the gain factor} to meet a convergence criterion which is, however, not performance relevant. \replaced[R1C04]{However, it requires the storage of the CSM, steering vectors, and beamforming maps for multiple frequencies in the memory. In terms of today's computational capacities, this should not be an issue, which }{The necessary memory scales linearly with the number of employed frequencies within the interval compared to standard CLEAN-SC, which in terms of today's computational capacities, should not be an issue. This }makes B-CLEAN-SC a viable method for little computational effort, but improved results at low and high frequencies.


\bibliography{main.bib}

\end{document}